\documentstyle[12pt]{article}
\begin{document}

\begin{flushright}
FERMILAB-Pub-96/120-A \\
May 1996
astro-ph/9605199
\end{flushright}

\vspace{1in}
\begin{center}
{\Large{\bf  Searching for stellar mass black holes in the solar neighborhood}}\\

\vspace{.4in}

{\bf Andrew F. Heckler and Edward W. Kolb} \\
{\em  NASA/Fermilab Astrophysics Center,
Fermi National Accelerator Laboratory,}\\
{\em Batavia, IL 60510, USA}\\{\em email: aheckler@fnas04.fnal.gov}
 
\vspace{.2in}
\begin{abstract}
We propose a strategy for searching for isolated stellar mass black holes in the solar neighborhood with the Sloan Digital Sky Survey. Due to spherical accretion of the inter-stellar medium and the ambient magnetic field, an isolated black hole is expected to emit a blended, thermal synchrotron spectrum with a roughly flat peak from the optical down to the far infra-red.  We find that  the Sloan Survey will be able to detect isolated black holes, in the considered mass range of 1--100$M_{\odot}$, out to a few hundred parsecs, depending on the local conditions of the ISM. We also find that the black holes are photmetrically distinguishable from field stars and they have a photometry similar to QSOs. They can be further singled out from QSO searches because they have a featureless spectrum with no emission lines. The Sloan Survey will likely find hundreds of  objects that meet these criteria, and to further reduce the number of candidates, we suggest other selection criteria  such as infra-red searches and proper motion measurements. Estimates indicate that dozens of black holes may exist out to a few hundred parsecs. If no black hole candidates are found in this survey, important limits can be placed on the local density of black holes and  the halo fraction in black holes, especially for masses greater than about $20 M_{\odot}$.

\end{abstract}
\end{center}

\vspace{.6in}
submitted to {\em Astrophysical Journal Letters}
\renewcommand{\thesection}{\Roman{section}} 

\newpage

Shvartsman (1971) first pointed out that a stellar mass black hole accreting matter from the interstellar medium (ISM) will have a flat, optical spectrum,  with no lines, and irregular fluctuations. Because of the expected rarity and  relatively weak luminosity of these objects, however,  a dedicated, systematic search has not been performed.  Nonetheless, Ipser and Price (1977) have placed limits on the halo population of large mass black holes ($M> 10^5 M_{\odot}$) based on non-detection in infrared surveys, and McDowell (1985) has argued that proper motion surveys have not revealed any obvious black hole candidates, and from this he has placed important (though model dependent) limits on both  the halo population  of black holes down to about $M= 10^3 M_{\odot}$, and the disk population down to about $M= 10 M_{\odot}$. Carr (1994) has also reviewed other limits placed on the density of black hole populations.

The imminent Sloan Digital Sky Survey (SDSS), however, offers an excellent opportunity to search for isolated black holes in our solar neighborhood in a more systematic manner. In fact, through planned QSO searches the SDSS will serendipitously collect both photometric and spectroscopic data on viable black hole candidates. Because ISM accretion on a black hole is a relatively simple problem, one can calculate the emergent spectrum and use it as a template to select candidates from the SDSS data. This systematic search can place much stronger limits on black hole populations if no candidates are found, especially when this survey is combined with proper motion studies and infrared surveys which will help to further single out black hole candidates.

The calculation of the emergent spectrum of a black hole spherically accreting the ISM was first done by Shvartsman (1971), and since then numerous authors have investigated the problem in detail (for example Zeldovich and Novikov 1971, Novikov and Thorne 1973, Shapiro 1973a, Ipser and Price (1982)).\footnote{For a recent review of black hole accretion, see Chakrabarti 1996.} The brems\-strahlung luminosity was found to be very weak, but if inter-stellar magnetic fields are included in the accretion process, the magnetic field is drawn in and compressed to higher field strengths ( about $10$ Tesla at the horizon for standard ISM conditions), and the resulting synchrotron luminosity can be quite high, with emission efficiencies  as large as $ 0.01 \dot{M}c^2$.

We have calculated numerically the emergent spectrum, following the  method of  Ipser and Price (1982). See Figure~1 for examples of expected spectra for various ISM densities and black hole masses. We only consider black holes with mass $M \le 100  M_{\odot}$ because the accretion rate for these ``smaller'' black holes is low enough that the accreting fluid can be considered optically thin for most ISM conditions. Thus, we will assume the accreting matter does not reabsorb any radiation that it emits. Because of the relative simplicity of the system, the calculation of the spectrum is fairly model independent, though there are still outstanding issues as to exactly how the  magnetic, gravitational, kinetic and thermal energies of the fluid are partitioned. Ipser and Price (1982) parameterize the amount of equipartioning in each form of energy (e.g., kinetic, magnetic), and for a reasonable ranges of  these parameters, the spectrum is not dramatically effected.

In our numerical calculations, we have added the effect of the black hole moving with velocity $V$. The effect of motion is easily accounted for by making the substitution (Bondi (1952))
\begin{equation}
a^{2}_{\infty} \rightarrow a^{2}_{\infty} +V^2.
\end{equation}
Note that $a_{\infty} \simeq (16.6{\rm km/sec})(T_{\infty}/10^4{\rm K})^{1/2}$ is the sound speed in the ISM with temperature $T_{\infty}$, where we have assumed that ISM is  polytropic with $P \propto n^{5/3}$.

Thus the total synchrotron luminosity of the black hole is 
\begin{eqnarray}\label{luminosity}
L_{\rm synch} &\approx & 4.5\times 10^{27} {\rm ergs \,\,s^{-1}} \beta_{d}\beta_{v}\left( \frac{M}{M_{\odot}} \right)^{3} \nonumber \\
& & \times\left( \frac{\rho_{\infty} }{10^{-24}{\rm g/cm}^3}\right)^{2}\left(\left( \frac{V}{16.6{\rm km/sec}} \right)^{2} + \frac{T_{\infty}}{10^4{\rm K}}  \right)^{-3},
\end{eqnarray}
where  $\beta_{v}$ and $\beta{d}$ are the equipartition parameters for bulk motion and dissipation, discussed by Ipser and Price (1982). We set  $\beta =  1$ unless otherwise stated. Note that eq.~(\ref{luminosity}) is valid for standard ISM conditions and for black holes of mass $M<100M_{\odot}$, where accretion rates are relatively small (Ipser and Price (1982)). As pointed out by Shvartsman (1971), the spectrum is  quite flat  and falls off exponentially above a cutoff frequency of $\nu_{\rm cutoff} \approx10^{15}$Hz for standard ISM conditions and $\nu_{\rm cutoff} $ is roughly independent of the black hole mass (see also Ipser and Price (1982)).

We should note here that this calculation of the emergent spectrum assumes the ISM is fully ionized. While it is true that as much as 70\% of the local ISM may be ionized (Cox and Reynolds 1987), it is is also true that in some regions it will be mostly neutral hydrogen. Strictly speaking, the fraction of ionization will effect the accretion rate and the temperature profile of the accreting fluid. For example, for a given $T_{\infty}$, a pure HI region will have a sound velocity smaller by a factor of $\sqrt{2}$ over a HII region, and this will increase the luminosity of the black hole. However, there are competing effects on the temperature profile. The in-falling HI gas will spend some of its gravitational energy on ionization, and this will result in lower temperatures at smaller radii, compared to the in-falling HII case. However, the lower temperature gas will undergo less synchrotron cooling, thus its temperature  will rise faster than in the HII case. Because there is a general equipartitioning of energy near the Schwartzschild radius $r_{s}$, where most of the emission takes place, one might expect the temperature profiles to be similar near $r_{s}$ in both the HI and HII cases, and the only significant difference between the two cases may be in the total luminosity and not emergent spectrum. A more precise answer warrants a  full calculation including  neutral hydrogen.

Now  that we have discussed the calculation of the emergent spectrum of the black hole accreting the ISM, let us consider a search strategy for black holes with the SDSS. The first step is to determine their apparent magnitude, taking into account their intrinsic luminosity and average expected distance. Using stellar evolution arguments, Shapiro and Teukolsky (1983) have estimated the number of stellar mass black holes in the galaxy to be  as much as $10^{8}$. Assuming that the black holes are homogeneously distributed in a halo of 20kpc, this translates to  a local number density of $3\times 10^{-6} $, and possibly more if the black holes are concentrated in the disk. Therefore, on average one would expect to find at least a few stellar mass black holes out to a distance of about $ 100$ parsecs. A similar calculation can be made by presuming that black holes comprise a significant fraction of the halo; this is discussed more below.

Whether or not a black hole in our solar neighborhood will be observed or not  will depend on the ISM in which it is imbedded. Unfortunately, we are in a hot, underdense local bubble of the ISM, and this significantly decreases the accretion rate, hence the luminosity, of the black holes. The general features of density and temperature of the local bubble have been mapped out (Cox and Reynolds 1987 and Diamond, Jewell and Ponman 1994). In the direction out of the galactic plane, the local bubble of very hot ($10^5$K) and very rare (0.05cm$^{-3}$) ISM is known to extend out as much as 200 parsecs. In the direction of the galactic center, however, the local bubble only extends out to a distance of about 60 parsecs. The local ISM is known to be inhomogeneous (Cox and Reynolds (1987)); for example, the sun is imbedded in a slightly over-dense region ($n\sim 0.1{\rm cm}^{-3}, T\approx 10^{3}$--$10^{4}K$) known as the ``local fluff,''  with a  size of about $ 20$parsecs.
Table~1 presents  the apparent magnitude of the black hole as a function of mass for several different values of density of the ISM. As both this table and eq.(\ref{luminosity}) show, the luminosity is a very sensitive function of  ISM conditions. 

\begin{table*}
\begin{center}
\begin{tabular}{r|ccc}
  &   \multicolumn{3}{c}{ $g$-magnitude}  \\
$M_{\rm BH}$ &($n = 1\,{\rm cm}^{-3}$) &($n = 0.1\,{\rm cm}^{-3}$) &(loc. bubble)\\
\hline
1 $M_{\odot}$ & 22.05&26.89 & 37.02\\
10 $M_{\odot}$ & 14.55 &19.39 & 29.52\\
100 $M_{\odot}$ & 7.38 &11.89 & 22.02
\end{tabular}
\end{center}
\caption{\normalsize{ The $g$-magnitudes of black holes at a distance of 10 pc,  for various black hole masses $M_{\rm BH}$ and  ISM densities $n$. For the local bubble, we set  $n=0.05\,{\rm cm}^{-3}$ and $T_{\rm ISM} = 10^5$K. Otherwise we set  $T_{\rm ISM} = 10^4$K, and for all cases $V=0$. To find magnitudes at other distances, add $5 \log{(D/10{\rm pc})}$ to the values in the table. The SDSS magnitude limit is about $m \simeq 22$.\label{tbl-1}}}
\end{table*}

As shown in eq.~(\ref{luminosity}), the velocity of the black hole is also an important factor in determining the luminosity if the black hole has a relative velocity much greater than the sound speed velocity of the ISM. Let us consider two populations of black holes: those co-rotating with the disk and those in the halo. If the black holes are co-rotating with the disk, then their average random velocity with respect to the disk will be about 10 km/sec, and for ISM temperatures greater than $10^{4}$K, the velocity will not be important. However, for the halo population of  black holes, the average random velocity is about 270 km s$^{-1}$, which is so large, that the luminosity is reduced by a factor of  about $10^{7}$ over the disk population case. As discussed later in this Letter, if we assume a Maxwellian velocity distribution, only a small fraction of halo black holes will have a relative velocity small enough to be sufficiently luminous. 

Assuming that there are black holes bright enough to be observed by the SDSS, we must then determine how distinguish them from the multitude of other sources such as field stars, galaxies and QSOs. Fortunately this may be not a difficult task. To demonstrate this, we have calculated the photometric colors (as defined by the AB magnitude system, which is used by the SDSS and is described in detail by Fukugita etal. 1996) of  a sample of several black holes,  and plotted them in Figure~2. The sample includes black holes ranging in mass from 3 to 100$M_{\odot}$ and imbedded in ISM with densities from 0.01 to 10 cm$^{-3}$. We also varied the equipartition parameter $\beta_{d}$ from 0.25 to 0.9. Note that changing the temperature of the ISM has a similar effect as changing the ISM density.  We have also plotted the colors of stars ($18.5<m_{r}<21.5$) and spectroscopically confirmed quasars taken from the data sample of  Trevese etal. (1994). The data has been transformed to SDSS colors (Newberg and Yanny 1995), and  overall errors are about 0.1 magnitudes.  This color-color plot reveals that black holes occupy only a small region of color space, and the black hole locus is easily distinguishable from the locus of main sequence stars. In addition, the black holes are in the same region of color space as QSOs, and inadvertently will thus be considered as QSO candidates. 

This is a fortuitous result for many reasons. The SDSS is expected to record about  $10^8$ objects, and the QSO search algorithm (Gunn 1995, Newberg and Yanny 1996) will cull this number down to about $10^{5}$  QSO (and hence black hole) candidates. In order to further single out black hole candidates, one can utilize the fact that spectra will be taken of all QSO candidates. The spectra will be helpful for two reasons. First of all, the black hole spectra will have no emission lines, so any spectra with lines can be omitted. Second, for a given black hole mass and for a given  ISM density and temperature, one can calculate the possible range of emergent spectra, and use these calculated spectra as a template for comparison with the measured spectra. Naturally, this is similar to saying that the spectra occupy a small region of color space.

Once one has found  a candidate that has no emission lines and has a spectrum similar to the black hole spectrum, how can one be sure that it is a black hole and not a star or some other object? For example, as pointed out by Shvartsman (1971), DC dwarfs may appear similar to accreting black holes.  Photometrically, DC dwarfs appear as quasars, and their spectra have no features or show  weak carbon features (Green, Schmidt and Liebert 1986). Novikov and Thorne also point out that neutron stars with little or no magnetic field may also have spectra similar to black holes. However, since a $1.4 M_{\odot}$ neutron star has a physical radius about three times larger than the same mass black hole, and since most of the radiation is emitted at small radii, it will probably not be difficult to distinguish between a black hole and a neutron star. For example, the boundary condition at the ``surface'' of the two objects are quite different, and this is precisely where most of the emission takes place. In addition, recall that synchrotron emission is the dominant mechanism, with fields approaching 10 Tesla. This field strength is probably small compared to even ``small'' magnetic field neutron stars, keeping in mind that pulsars have magnetic fields of about $10^9$ Tesla.

If black hole candidates are found with the SDSS, then there are still several additional tests one can perform in order to be more confident that the candidate is a nearby accreting black hole. Perhaps the most detailed test will involve observations of the spectrum at longer wavelengths. As shown explicitly by Ipser and Price (1982), the spectrum is mostly flat and extends down to a  lower  cutoff frequency $\nu_{\rm low} \approx 10^{12}$ -- $10^{13}$Hz, depending on the balck hole mass in ISM conditions, and below this frequency, the spectrum goes as $\nu^{2}$. Thus the spectrum can extend well into the low end  of the infrared. This suggests that  one may be able to search for the infrared signal of the black hole candidates with infrared surveys such as the IRAS point source survey. To give an example, for a $30 M_{\odot}$ black hole in standard ISM conditions at a distance of 10 pc, $m_{u} \approx  12$ and the expected flux at $30\mu$ is of order 0.1Jy, which is on the lower end of what is measurable by IRAS.

Besides the peculiar flat spectrum of the black hole, one can also test for variability in the luminosity and spectrum. Novikov and Thorn (1973) have estimated that instabilities in the plasma will result in variability in the luminosity on times scales of $10^{-4}$ -- $10^{-2}$ sec. In addition,  they point out that there will be variability due to the black hole passing through different regions of the inhomogeneous ISM. They estimate the minimum time scales for this variability to be 
\begin{equation}\label{}
\Delta t > 2b_{\rm capture}/V \approx 10\,{\rm years} \left( \frac{M}{M_{
\odot}} \right)\left( \frac{V}{10{\rm km/sec}} \right)^{-3} .
\end{equation}
Naturally, the variability is not expected  to be periodic.

As pointed out by McDowell (1985), since the black holes are so close, another important test is to measure their proper motion and distance, which may be measured by parallax. Even if the black hole is farther than 100 pc away and has a tangential velocity of 10km/sec, one should be able to measure the proper motion over the span of a few years.  One can then place an important constraint on black hole candidates by verifying whether the velocity and distance measurements correspond with the luminosity derived from  information on the density and temperature of the ISM and a reasonable range of black hole masses. Unfortunately, because the shape of the black hole spectra for different mass is so similar, it will be difficult to determine the mass of the black hole if the distance and/or ISM conditions are not known. 

One important question is how many objects one should expect to be selected as black  hole candidates in the Sloan Survey? We estimate this by noting that in the Palomar-Green catalog of ultraviolet-excess stellar objects by Green, Schmidt and Liebert (1986), about $0.3\%$ of the objects are unknown, that is they have no identified spectral features. If we assume that the SDSS will see approximately $10^{5}$ UV excess objects, then we roughly estimate that one will find about $300$ unknown UV excess objects. Since accreting black holes have no spectral features, then we identify these unknown objects as black hole candidates. The number of candidates may be reduced even further by selecting only those featureless objects which have photometry similar to black holes.

Finally, let us make one more important point. Because of the relatively high luminosity of black holes with mass $M > 10 M_{\odot}$, there are important consequences if no black holes are found in the Sloan Survey.  To illustrate this let us determine the maximum amount black holes can contribute to the dark matter halo, if they are not detected in SDSS.   The number of black holes $N_{\rm obs}$ observable by SDSS is
\begin{eqnarray}\label{}
N_{\rm obs} = n_{\rm BH}\left( \frac{M_{\odot}}{M} \right) \Omega\left(\frac{d^{3}_{\rm max}(M,\rho_{\infty},T_{\infty})}{3}  \right)  \eta f(M)
\end{eqnarray}
where $n_{\rm BH}\approx 0.01 M_{\odot}/{\rm pc}^3$ is the local dark matter halo density (Gates, Gyuk and Turner (1995)), $\Omega$ is the solid angle of sky coverage, $d_{\rm max}$ is the maximum distance to which the SDSS can observe the spectrum of a black hole, given that it has a limiting magnitude of $m_{r} =20$, and  $f(M)$ is the fraction of the halo with black holes of mass $M$. The factor $\eta$ accounts for the fact that the black holes in the halo will have a high velocity with respect to the ISM. The luminosity is a very sensitive function of velocity (see eq.(\ref{luminosity})), thus we include only those black holes with  velocity $\bf{v}$ such that $|{\bf v} - {\bf V}| < a_{\infty}$, where $\bf{V}$ is the local velocity of the disk. We use the simple model of a non-rotating halo which has a locally isotropic and Maxwellian distribution with an average velocity of 270 km/sec. $\eta$ is the fraction of the population which obeys $|{\bf v} - {\bf V}| < a_{\infty}$. For $T_{\infty} = 10^{4} K$,  $\eta \approx 10^{-4}$. By setting $N_{\rm obs} \le 1$, using $\Omega = \pi$, and assuming 100\% detection efficiency, we obtain the result for $f(M)$, shown in Figure~3. Note that the SDSS will be very sensitive to black holes of mass $M {\ \lower-1.2pt\vbox{\hbox{\rlap{$>$}\lower5pt\vbox{\hbox{$\sim$}}}}\ } 10 M_{\odot}$.

We would especially like to thank Brian Yanny, Heidi Newberg, and  Rich Kron for providing the data on the star and QSO colors, and for providing useful discussions on both the data and on the SDSS QSO search program. This work was supported in part by the DOE and by NASA (NAG5-2788) at Fermilab.

\vspace{.15in}
\noindent {\Large{\bf References}}
\vspace{.15in}

\noindent  Bondi, H. 1952, {\em MNRAS}, {\bf 112}, 195.

\noindent Carr, B. 1994, {\em A.A.R\&A.}, {\bf 32}, 531.

\noindent Chakrabarti, S.K. 1996, {\em Phys. Rep.}, {\bf 266}, 229.

\noindent Cox, D. and Reynolds, R. 1987, {\em A.A.R\&A.}, {\bf 28}, 215.

\noindent  Diamond, C. J., Jewell, S.J. and Ponman, T.J. 1995, {\em MNRAS}, {\bf 274}, 589.

\noindent Fukugita, M. etal. 1996, {\em AJ}, {\bf 111}, 1748.

\noindent Gates, E.I., Gyuk, G., and Turner, M.S. 1995, {\em Ap.J.}, {\bf 449}, L123.

\noindent Green, R.F. , Schmidt, M. and Liebert, J. 1986, {\em Ap. J. Supp.}, {\bf 61}, 305.

\noindent Gunn, J. E.  1995, Bull. American Astron. Soc., {\bf 186}, \#44.05.

\noindent Ipser, J. R. and Price, R. H. 1977, {\em Ap.J.}, {\bf 216}, 578.

\noindent Ipser, J. R. and Price, R. H. 1982, {\em Ap.J.}, {\bf 255}, 654.

\noindent  McDowell, J. 1985, {\em MNRAS}, {\bf 217}, 77.

\noindent Newberg, H. and Yanny, B. 1995, Fermilab Tech. Memo. TM-1973.

\noindent Novikov, I. D. and Thorne, K. S. 1973, in {\em Black Holes}, ed. C. DeWitt and B. DeWitt (New York: Gordon \& Breach).

\noindent Shapiro, S. L. $1973a$, {\em Ap.J.}, {\bf 180}, 531.


\noindent Shapiro, S.L. and Teukolsky, S.A., 1983 {\em Black Holes, White Dwarfs and Neutron Stars.} Wiley, New York.

\noindent  Shvartsman, V. F. 1971, {\em Soviet Astr.--AJ},{\bf 15}, 377.

\noindent Tevese, D., Kron, R.G., Majewski, S.R. and Koo, D.C. 1994 {\em Ap.J.}, {\bf 433}, 494.

\noindent Zel'dovich, Ya. B. and Novikov, I. D. 1971, {\em Relativistic Astrophysics}, Vol.~{\bf 1} (Chicago: University of Chicago Press).

\listoffigures

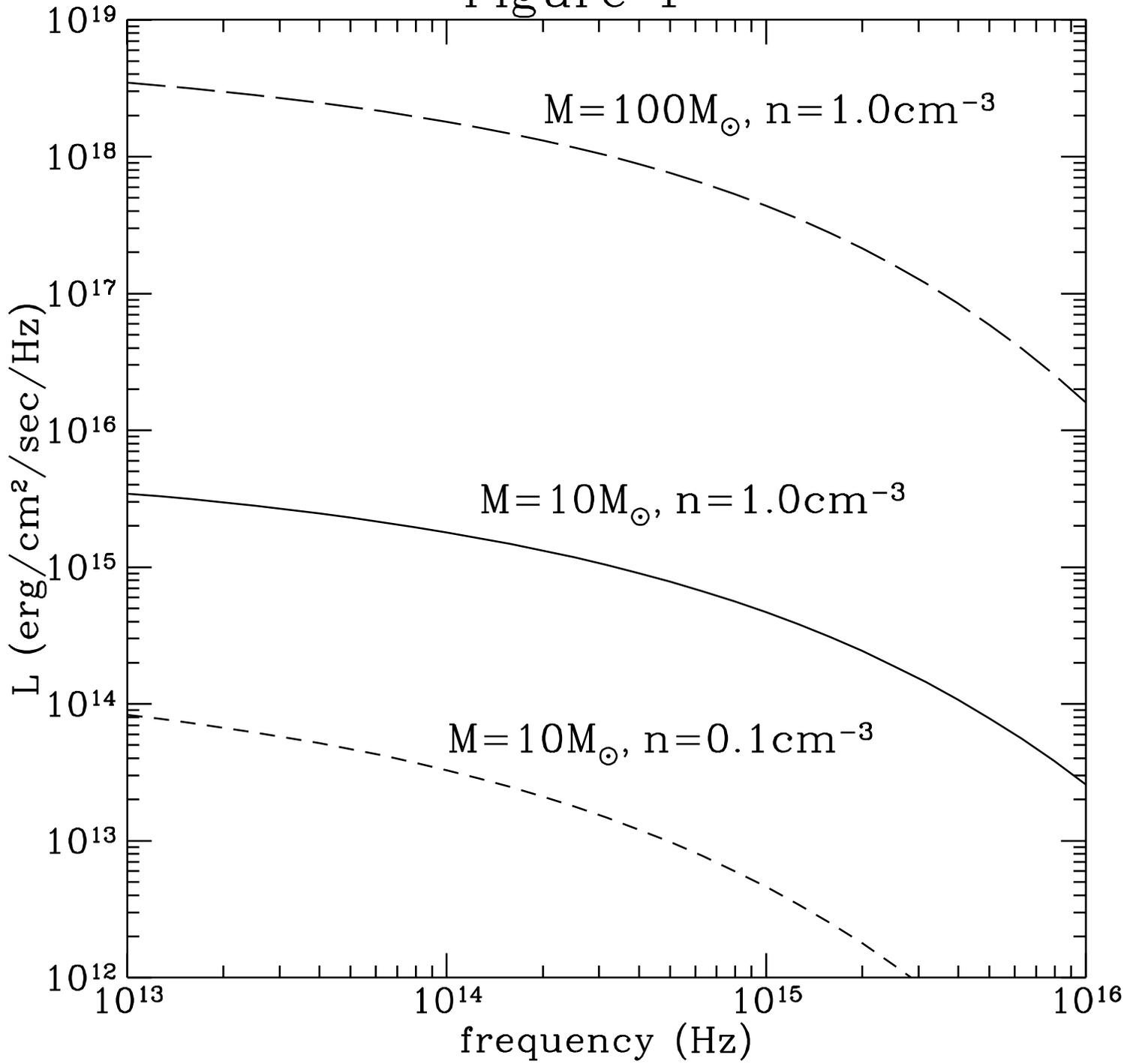
Figure1. Emergent spectra of accreting black holes for various black hole masses and ISM densities.

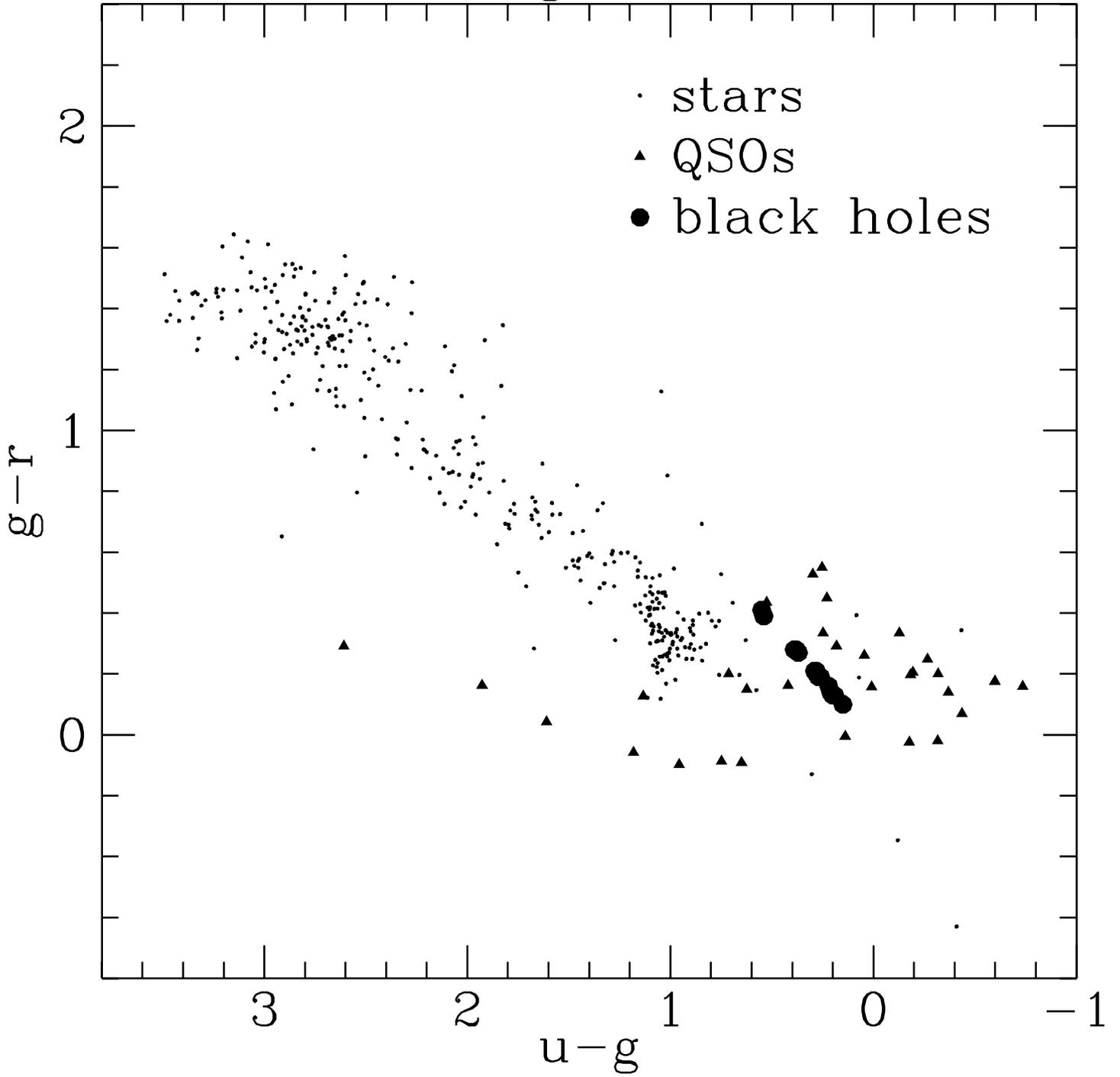
Figure 2. Color-color diagram for stars and QSOs, taken from Trevese etal. (1994), and the expected colors of isolated black holes accretting the ISM. Notice that the black hole locus is distinct from the main-sequence stellar locus, and lies within the QSO locus. The SDSS QSO search will consider objects ``bluer'' (below and to the right) than the main-sequence stars to be QSO candidates.

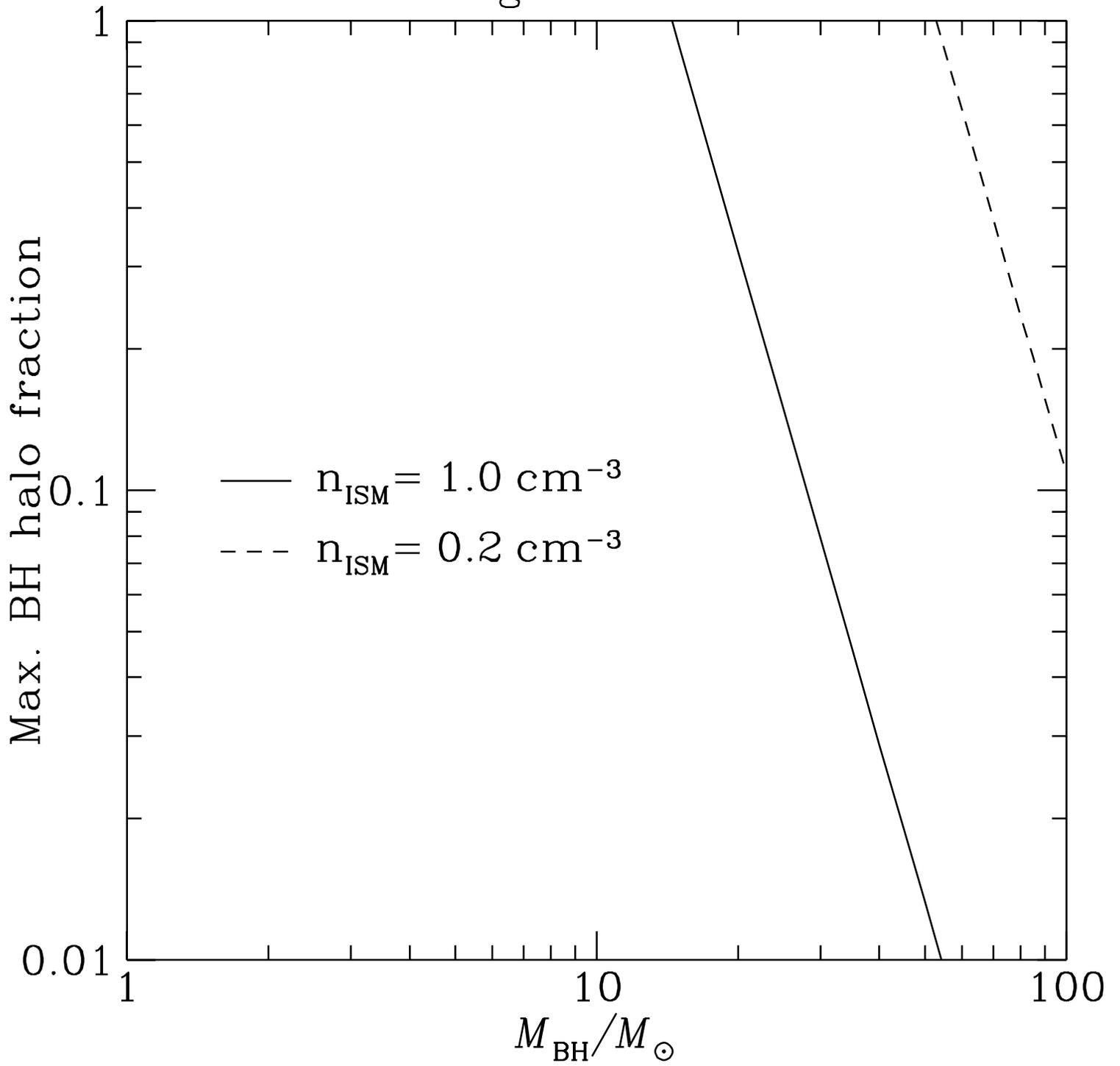
Figure 3. Maximum halo mass fraction in black holes if no black hole candidates are found with the SDSS, for two values of the ISM density. For $M{\ \lower-1.2pt\vbox{\hbox{\rlap{$>$}\lower5pt\vbox{\hbox{$\sim$}}}}\ } 10 M_{\odot}$, the SDSS is sensitive to black holes beyond the local bubble, where one expects $\bar{n}_{\rm ISM} \sim 1\,{\rm cm}^{-3}$.

\end{document}